\documentclass[a4paper,runningheads]{llncs}

\usepackage[T1]{fontenc}
\usepackage{amsmath,amssymb,amsfonts}
\usepackage{newtxtext}
\usepackage[varvw]{newtxmath}
\usepackage{graphicx}
\usepackage{float}
\usepackage{booktabs}
\usepackage{longtable}
\usepackage[pass]{geometry}
\usepackage{array}
\usepackage{multirow}
\usepackage{microtype}
\usepackage[section]{placeins}
\usepackage{siunitx}
\usepackage{tikz}
\usetikzlibrary{arrows.meta,backgrounds,calc,fit,positioning,shapes.geometric}
\usepackage{hyperref}
\usepackage{macro}

\title{\sys: Towards a Programmable Strategy Interface for Flexible SAT Solving}
\titlerunning{\sys: Towards a Programmable Strategy Interface for Flexible SAT Solving}
\author{Haoze Wu\inst{1} \and Andr\'{e} Schidler\inst{3} \and
Armin Biere\inst{3} \and Nina Narodytska\inst{2}}
\authorrunning{H. Wu et al.}
\institute{Amherst College, Amherst, MA, USA
\and Broadcom, Palo Alto, CA, USA
\and University of Freiburg, Freiburg im Breisgau, Germany}

\begin{document}
\maketitle
\begin{abstract}
Modern SAT solvers combine a rich repertoire of proof rules with sophisticated strategies that determine when and how to apply them. Often, the internal implementations are tightly coupled to particular strategies, making it inconvenient to define new strategies without changing solver internals. We present \sys, a proof-producing sequential CDCL solver that incorporates a wide range of techniques in state-of-the-art solvers with a clean separation of proof rules from solving strategies. This separation enables a public interface through which users can invoke particular techniques, observe their effects, and program a wide range of solving strategies. The default strategy of \sys is programmable over this interface and outperforms state-of-the-art solvers on benchmarks from the 2024--2026 SAT Competitions. We also use the interface to develop novel strategies tailored to challenging benchmarks from two application domains, which led to performance gains over strong baselines. We intend \sys to serve as an infrastructure for designing and evaluating novel  solving strategies. Finally, we also present a self-study of the AI-assisted development process behind Tactix.

\par\smallskip
\noindent \textbf{Code:} {\color{purpleheart} \hyperref[github.com/ai-ar-research/Tactix/]{https://github.com/ai-ar-research/Tactix/}}
\end{abstract}

\section{Introduction}
\label{sec:intro}

Modern SAT solvers achieve impressive practical performance by carefully orchestrating a rich set of techniques, including conflict-driven clause learning, restarts, local search, and various inprocessing rules such as bounded variable elimination~\cite{een2005preprocessing}, clause vivification~\cite{piette2008vivifying}, and bounded variable addition~\cite{manthey2012reencoding}.
As the repertoire of reasoning techniques expands, the number of possible ways to combine and apply them grows combinatorially, making it increasingly challenging to identify effective solving strategies. Modern SAT solvers typically employ hard-coded strategies that are carefully tuned on a broad range of benchmarks, with their internal implementations tightly coupled to these strategies to maximize performance. Although solvers often expose configuration options for adjusting aspects of their strategies (e.g., whether to enable particular inprocessing rules or how much computational effort to allocate to a technique), these options still cover only a limited subset of the possible strategies and their combinations, potentially excluding those better suited to specific problem domains of interest. Consequently, in order to experiment with a new solving strategy that significantly deviates from the default, the user often has to introduce changes to the solver's internal code and understand the solver's internal implementation well enough to ensure that their modification does not introduce unintended correctness or performance issues.

Ideally, a SAT solver should cleanly separate the implementation of solving techniques from the strategies that orchestrate their application. Such a separation would allow users to program and experiment with a rich range of strategies without modifying the solver's internals, potentially making more effective use of the available techniques. It would also facilitate a systematic study of solver behavior and provide a unified framework in which different SAT solvers can be expressed as distinct strategies over a common set of techniques. The separation of proof rules from solving strategies is well established in abstract calculi~\cite{nieuwenhuis2006abstract}, and its practical appeal has also long been recognized~\cite{maric2009formalization,ganeshfmsd22}. However, realizing this vision in a practical, high-performance SAT solver raises several non-trivial design questions: at what granularity should the interface expose individual techniques? For example, should users invoke bounded variable elimination on one variable at a time, or should they be able to perform a batch of eliminations governed by a low-level variable selection heuristic and a computational budget? How can we strike a balance between maintaining the efficiency of the implementation and allowing the application of a rich set of techniques at arbitrary stages of the solving process? How can diverse techniques from different solvers be incorporated into a uniform backend?

In this paper, we take a step towards such rule/strategy separation in SAT-solving and introduce \sys, a proof-producing sequential CDCL solver with a public interface for defining solving strategies. This interface allows a user to implement a strategy as a client program that selects among the currently available operations, specifies how they should be applied, observes their effects, and decides which operation to invoke next. 
For example, the strategy may ask to perform bounded variable elimination with specific limits on clause growth and the number of resolution attempts. The \sys backend would perform the operation, and the strategy can then inspect why the operation stopped (e.g., completed, budget reached, or encountered a conflict) and the effect of the operation (e.g., how many resolution attempts were performed and how many variables were eliminated). Based on this information, the strategy may repeat the operation or select another operation. Throughout this process, the solver core maintains the formula and search state, emits any proof steps required by a transformation, and records the information needed to reconstruct a satisfying assignment to the original formula. \sys incorporates a wide range of search, simplification, symmetry-breaking, and structural reasoning techniques from existing tools, including \kissat~\cite{biere2021kissat}, \cadical~\cite{biere2024cadical}, \satsuma~\cite{anders2024satsuma}, \cryptominisat~\cite{soos2009cryptominisat}, and \lymphosat~\cite{green2026hyperspecialization}, and, depending which techniques was invoked, generates UNSAT proofs in DRAT~\cite{wetzler2014drat}, DSR~\cite{codel2024verified}, or PBP~\cite{bogaerts2023certified} formats.

\sys also comes with a default strategy that is programmable over the public interface. This strategy employs the interval/effort-based rule-scheduling employed in \kissat. On benchmarks from the 2024--2026 SAT Competitions, this default strategy performs competitively against winners of the 2024--2026 SAT Competitions, solving more instances overall and contributing unique solutions. To further understand the usefulness of the public interface, we co-design with an AI agent new solving strategies over the public interface for two sets of challenging benchmarks, one for arithmetic circuit verification~\cite{ritirc2017column,kaufmann2019arithmetic}, the other for tackling finite cases of the Chv\'atal's conjecture~\cite{chvatal_conjecture_1974,looney2026partitioning}. In both cases, we were able to come up with a novel strategy that deviates significantly from canonical strategies and achieves substantial gains over baseline solvers in terms of solving and proof-checking time.

In addition to the system description of \sys, we also describe our solver development process, including how we perform correctness and performance testing, define milestones, and leverage AI agents. We reflect on the experience of developing \sys and discuss infrastructures that are currently under-developed in the formal methods community that can facilitate co-developing solvers with AI agents in a more efficient and systematic manner.

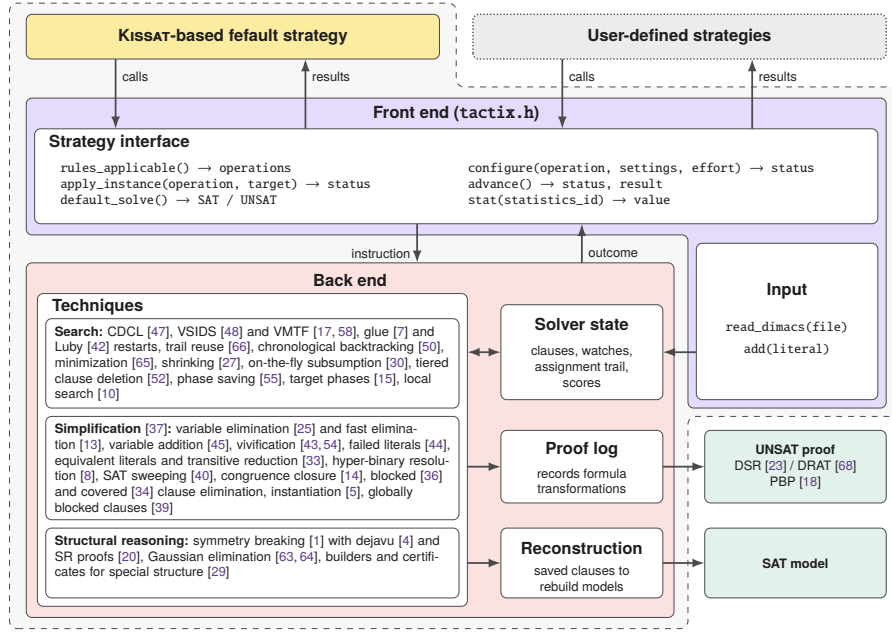
\begin{figure}[t]
  \centering
  \resizebox{\linewidth}{!}{
\definecolor{stratfill}{HTML}{FCEC9F}
\definecolor{ifacefill}{HTML}{E4DBFD}
\definecolor{kernfill}{HTML}{FBE1E0}
\definecolor{outfill}{HTML}{E3F1EC}
\definecolor{ink}{HTML}{1E1E1E}
\definecolor{edge}{HTML}{4A4A4A}
\begin{tikzpicture}[x=1cm,y=1cm,font=\sffamily,
  box/.style={draw=edge,line width=0.5pt,rounded corners=3pt},
  head/.style={font=\sffamily\bfseries\small,text=ink},
  sub/.style={font=\sffamily\bfseries\footnotesize,text=ink},
  lab/.style={font=\sffamily\scriptsize,text=ink,align=center},
  txt/.style={font=\sffamily\scriptsize,text=ink,align=left,anchor=north west},
  api/.style={font=\ttfamily\scriptsize,text=ink,align=center},
  arr/.style={-{Latex[length=2.2mm,width=1.6mm]},line width=0.9pt,draw=edge},
  sarr/.style={-{Latex[length=1.8mm,width=1.3mm]},line width=0.7pt,draw=edge},
  tag/.style={font=\sffamily\scriptsize,inner sep=1pt,text=ink}]

  \draw[edge,dashed,line width=0.5pt,fill=gray!6,rounded corners=4pt]
    (-0.3,-0.2) -- (11.85,-0.2) -- (11.85,3.6) -- (15.65,3.6) -- (15.65,9.5)
    -- (7.7,9.5) -- (7.7,11) -- (-0.3,11) -- cycle;

  \draw[box,fill=stratfill] (0,10.0) rectangle (7.4,10.8);
  \node[head] at (3.7,10.4) {\kissat{}-based fefault strategy};
  \draw[draw=edge,line width=0.7pt,rounded corners=3pt,densely dotted,fill=gray!15] (8.0,10.0) rectangle (15.4,10.8);
  \node[head] at (11.7,10.4) {User-defined strategies};

  \draw[box,fill=ifacefill] (0,6.85) -- (11.85,6.85) -- (11.85,3.75) -- (15.4,3.75)
    -- (15.4,9.3) -- (0,9.3) -- cycle;
  \node[head] at (7.7,9.03) {Front end (\texttt{tactix.h})};
  \draw[box,fill=white] (0.15,7.05) rectangle (15.25,8.75);
  \node[sub,anchor=west] at (0.3,8.5) {Strategy interface};
  \node[font=\ttfamily\scriptsize,text=ink,align=left,anchor=north west] at (0.5,8.25) {%
    rules\_applicable() $\to$ operations\\
    apply\_instance(operation, target) $\to$ status\\
    default\_solve() $\to$ SAT / UNSAT};
  \node[font=\ttfamily\scriptsize,text=ink,align=left,anchor=north west] at (7.8,8.25) {%
    configure(operation, settings, effort) $\to$ status\\
    advance() $\to$ status, result\\
    stat(statistics\_id) $\to$ value};
  \draw[box,fill=white] (12.0,3.9) rectangle (15.25,6.7);
  \node[sub] at (13.625,5.85) {Input};
  \node[api] at (13.625,5.0) {read\_dimacs(file)\\[3pt]add(literal)};

  \draw[box,fill=kernfill] (0,0) rectangle (11.6,6.35);
  \node[head] at (5.8,6.05) {Back end};
  \draw[box,fill=white] (0.2,0.2) rectangle (7.9,5.8);
  \node[sub,anchor=west] at (0.35,5.57) {Techniques};
  \draw[box,fill=white] (0.35,3.75) rectangle (7.75,5.35);
  \node[txt,text width=7.05cm] at (0.4,5.33) {\textbf{Search:} CDCL~\cite{marquessilva1999grasp},
    VSIDS~\cite{moskewicz2001chaff} and VMTF~\cite{ryan2004thesis,biere2015vmtf}, glue~\cite{audemard2009glucose}
    and Luby~\cite{luby1993optimal} restarts, trail reuse~\cite{vandertak2011reusing}, chronological
    backtracking~\cite{nadel2018chrono}, minimization~\cite{sorensson2009minimizing},
    shrinking~\cite{fleury2021shrinking}, on-the-fly subsumption~\cite{han2009otfs}, tiered clause
    deletion~\cite{oh2015between}, phase saving~\cite{pipatsrisawat2007phase}, target
    phases~\cite{biere2020kissat}, local search~\cite{balint2012probsat}};
  \draw[box,fill=white] (0.35,1.75) rectangle (7.75,3.6);
  \node[txt,text width=7.05cm] at (0.4,3.58) {\textbf{Simplification}~\cite{jarvisalo2012inprocessing}\textbf{:}
    variable elimination~\cite{een2005preprocessing} and fast elimination~\cite{biere2024kissatsc},
    variable addition~\cite{manthey2012reencoding}, vivification~\cite{piette2008vivifying,luo2017vivification},
    failed literals~\cite{lynce2003probing}, equivalent literals and transitive
    reduction~\cite{heule2011bin}, hyper-binary resolution~\cite{bacchus2003hyper}, SAT
    sweeping~\cite{kuehlmann2002robust}, congruence closure~\cite{biere2024congruence}, blocked~\cite{jarvisalo2010bce}
    and covered~\cite{heule2010cce} clause elimination, instantiation~\cite{andersson2002proof},
    globally blocked clauses~\cite{kiesl2019truth}};
  \draw[box,fill=white] (0.35,0.35) rectangle (7.75,1.6);
  \node[txt,text width=7.05cm] at (0.4,1.58) {\textbf{Structural reasoning:} symmetry
    breaking~\cite{anders2024satsuma} with dejavu~\cite{anders2021dejavu} and SR
    proofs~\cite{buss2021sr}, Gaussian elimination~\cite{soos2009cryptominisat,soos2020tinted},
    builders and certificates for special structure~\cite{green2026hyperspecialization}};

  \draw[box,fill=white] (8.5,3.9) rectangle (11.4,5.6);
  \node[sub] at (9.95,5.25) {Solver state};
  \node[lab] at (9.95,4.5) {clauses, watches,\\assignment trail,\\scores};
  \draw[box,fill=white] (8.5,2.05) rectangle (11.4,3.35);
  \node[sub] at (9.95,3.0) {Proof log};
  \node[lab] at (9.95,2.45) {records formula\\transformations};
  \draw[box,fill=white] (8.5,0.35) rectangle (11.4,1.6);
  \node[sub] at (9.95,1.25) {Reconstruction};
  \node[lab] at (9.95,0.7) {saved clauses to\\rebuild models};

  \draw[box,fill=outfill] (12.15,2.05) rectangle (15.4,3.35);
  \node[lab] at (13.775,2.7) {\textbf{UNSAT proof}\\DSR~\cite{codel2024verified} / DRAT~\cite{wetzler2014drat}\\PBP~\cite{bogaerts2023certified}};
  \draw[box,fill=outfill] (12.15,0.35) rectangle (15.4,1.6);
  \node[lab] at (13.775,0.975) {\textbf{SAT model}};

  \draw[arr] (1.6,10.0) -- node[tag,right=2pt,pos=0.25]{calls} (1.6,8.75);
  \draw[arr] (5.0,8.75) -- node[tag,right=2pt,pos=0.75]{results} (5.0,10.0);
  \draw[arr] (9.6,10.0) -- node[tag,right=2pt,pos=0.25]{calls} (9.6,8.75);
  \draw[arr] (13.0,8.75) -- node[tag,right=2pt,pos=0.75]{results} (13.0,10.0);
  \draw[arr] (7.0,7.05) -- node[tag,left=2pt,pos=0.75]{instruction} (7.0,6.35);
  \draw[arr] (9.95,6.35) -- node[tag,right=2pt,pos=0.25]{outcome} (9.95,7.05);
  \draw[arr,{Latex[length=2.2mm,width=1.6mm]}-{Latex[length=2.2mm,width=1.6mm]}] (7.9,4.75) -- (8.5,4.75);
  \draw[arr] (7.9,2.7) -- (8.5,2.7);
  \draw[arr] (7.9,0.975) -- (8.5,0.975);
  \draw[arr] (12.0,4.75) -- (11.4,4.75);
  \draw[arr] (11.4,2.7) -- (12.15,2.7);
  \draw[arr] (11.4,0.975) -- (12.15,0.975);
\end{tikzpicture}}
  \caption{Tactix system overview with selected interface functions.}
  \label{fig:architecture}
\end{figure}

The rest of the paper is organized as follows. Section~\ref{sec:architecture} represents the bulk of the contributions and describes the design of \sys, including the backend and the public interface for defining solving strategies. Section~\ref{sec:harness} describes our AI-assisted software development process. Section~\ref{sec:evaluation} presents an empirical evaluation of \sys on SAT competition benchmarks, a set of arithmetic verification benchmarks, and subproblems of Chv\'atal's conjecture. Section~\ref{sec:related} reviews related work. We conclude and discuss future directions in Section~\ref{sec:conclusion}.

\section{\sys: A SAT Solver with a Programmable Strategy Interface}
\label{sec:architecture}

This section introduces the overall architecture of \sys, describes the operations supported by the backend, and explains how the public interface allows users to compose them into strategies.

\subsection{System overview}

Figure~\ref{fig:architecture} shows a system overview of \sys. \sys can be used as an off-the-shelf SAT solver or as a library where a user can create their own solving strategy. The executable reads a formula in DIMACS format and solves it with a default strategy. A library client creates a solver instance, supplies the formula through the input functions of the C interface (reading a DIMACS file or adding clauses), and can either invoke the default strategy or define their own strategies. The dashed boundary encloses the shipped codebase.

The frontend, \texttt{tactix.h}, is the public interface. It consists of the input and strategy interfaces, which provide the building blocks for programming solving strategies, including methods for invoking operations, observing their effects (e.g., number of forced literals), and inspecting solver states. The default strategy of \sys, which is based on that of \kissat, can be programmed directly through this public interface.\footnote{The same strategy is also implemented by calling the same operations in the C++ backend, which is used by the compiled \sys executable. Empirically (Section~\ref{sec:evaluation}), the two implementations yield identical behaviors, and the API version only introduces minor performance overhead.}
Each operation can specify its own settings and computational budget (a.k.a. effort). The core applies the selected technique, maintains the solver state, and records proof steps and model-reconstruction data as required. The public interface can also inspect the outcome of the application, which can help decide the next move.
The backend implements the operations and maintains the formula, search state, proof, and model-reconstruction data. To support different operation sequences, it checks the conditions required by each technique: for example, unit propagation must converge before a new branching decision is made, and bounded variable elimination can only be applied at decision level zero. An accepted request executes the selected operation and reports the outcome and effects. 

The solver produces either SAT or UNSAT. A SAT result comes with a satisfying assignment to the original input variables. When proof production is enabled, an UNSAT result comes with a certificate of unsatisfiability that can be checked externally by a proof checker. The main proof format is deletion substitution redundancy (DSR), which is a superset of the DRAT format~\cite{wetzler2014drat} and supports ordinary clausal deductions together with satisfiability-preserving restrictions such as symmetry breaking~\cite{codel2024verified,buss2021sr}. \sys can also sometimes produce pseudo-Boolean proof~\cite{bogaerts2023certified}, as discussed later in this section.

\subsection{Solver backend}
\label{sec:action-contract}

\sys's back end integrates CDCL search, simplification, and structural reasoning. Simplification transforms the formula before and between periods of search~\cite{een2005preprocessing,jarvisalo2012inprocessing}. Structural reasoning exploits relationships such as symmetries~\cite{anders2024satsuma} and systems of equations~\cite{soos2009cryptominisat} represented by the clauses to derive new clauses or simplify the formula. When the back end receives a request to perform an operation, it checks the applicability of the operation, updates the shared solver state, performs the bookkeeping needed to keep the data structures consistent for subsequent operations, and records the information needed to certify its result. This organization allows a strategy to combine techniques while retaining the progress made by earlier operations. The full list of registered operations is available in the source distribution.\footnote{\texttt{src/rules/registry.cpp}}

\subsubsection{Search and shared state.} \sys implements a standard CDCL search procedure~\cite{marquessilva1999grasp}. It stores the current clauses in one clause database and the sequence of assignments in a \emph{trail}. Each propagated assignment has a \emph{reason}: the clause that forced it. Conflict analysis follows these reasons to derive a learned clause. Assignments are grouped by decision level, which counts the branching decisions currently in effect; level zero contains assignments derived without branching. These structures are shared by all techniques. For example, a clause derived by algebraic reasoning~\cite{soos2009cryptominisat} can force an assignment by unit propagation and subsequently serve as a reason during conflict analysis.
\sys incorporates a number of modern features of CDCL search. These include the VSIDS~\cite{moskewicz2001chaff} and variable-move-to-front (VMTF)~\cite{biere2015vmtf} heuristics for branching, \glucose{}-style and Luby-style restarts~\cite{biere2020kissat,luby1993optimal}, tiered learned-clause deletion~\cite{oh2015between}, and different phase saving techniques~\cite{pipatsrisawat2007phase,biere2020kissat}. \sys also contains a local search module that can be invoked with its own effort limit and search heuristics. While the front end allows the users to define search behavior at fine granularity (e.g., deciding on specific variables rather than using the pre-defined branching heuristics), these pre-defined mechanisms are available as useful building blocks for defining solving strategies.

\subsubsection{Simplification.}
Tactix supports a number of formula simplification techniques from \kissat and \cadical, such as subsumption~\cite{biere2021kissat,biere2024cadical} and vivification~\cite{piette2008vivifying} for clause reduction and simplification, failed-literal probing~\cite{lynce2003probing} for look-ahead style variable fixing, congruence closure~\cite{biere2024congruence} and SAT sweeping~\cite{kuehlmann2002robust} for discovering equivalences and forced literals, substitution of equivalent literals~\cite{heule2011bin}, bounded variable elimination~\cite{een2005preprocessing}, and bounded variable addition that introduces fresh variables to factor repeated clause structure and reduce the number of clauses~\cite{manthey2012reencoding}.
Similar to how search is implemented, \sys incorporates pre-defined low-level heuristics for each technique. For variable elimination, for example, it prioritizes variables using estimates of clause growth and revisits its choices as the formula changes, following \kissat{}'s elimination heuristics~\cite{biere2024kissatsc}. While a user could, for example, select individual variables to eliminate or clauses to vivify, it is often reasonable to leave these low-level choices to pre-defined heuristics.

\subsubsection{Structural reasoning.}
In addition to the aforementioned simplification techniques, \sys also incorporates techniques that discover specific structures in the formula to derive consequences or guide the search for a model. Specifically, \sys incorporates a number of symmetry-breaking techniques including orbitopal fixing, clausal fixing, and negation fixing~\cite{anders2026orbitopal,anders2024satsuma}. It also incorporates Schreier--Sims table cuts, which add binary clauses ordering symmetric literals~\cite{anders2026simplify}. Symmetry detection relies on the dejavu graph-automorphism library~\cite{anders2021dejavu}, applied to a graph representation of the current formula. Instead of directly invoking a mature tool like \satsuma, we implement these techniques within \sys's back end to support strategies that apply symmetry breaking between CDCL searches. In addition, \sys detects systems of equations over GF(2) and applies the Gaussian-elimination approach used in \cryptominisat{}~\cite{soos2009cryptominisat,soos2020tinted}, translating deductions back into clauses that search can use.
\sys also incorporates selected techniques from \lymphosat{}~\cite{green2026hyperspecialization}, a \satzilla{}-style~\cite{xu2008satzilla} solver that selects among over 120 specialized solvers. We prioritize techniques that focus on common algebraic and combinatorial structures, such as parity equations, exact-cover constraints, and independent sets, and generalize methods that rely on hard-coded hints. Concretely, we incorporated twelve model-construction-only procedures as local search and seven proof-producing refutation procedures for structures such as parity, cardinality, and Ramsey constraints. Each procedure recognizes constraints encoded by the clauses and operates on the recovered structure, translating its result back into a Boolean variable assignment or a refutation proof. All these additions can be interleaved with search and simplification as they operate on the same backend state.

\subsubsection{Proof production.} For UNSAT instances, \sys by default produces DRAT proofs~\cite{wetzler2014drat}, including for GF(2) reasoning~\cite{philipp2016drat}. These are extended to DSR (a superset of DRAT) when techniques such as symmetry breaking require substitution-redundancy steps~\cite{buss2021sr,codel2024verified}. The backend checks each substitution-redundancy step before applying it, either against the current formula or leveraging previously verified structural properties such as symmetries. \sys contains two preprocessing procedures derived from \lymphosat{} that produce pseudo-Boolean (PBP~\cite{bogaerts2023certified}) proofs: one derives contradictions by summing XOR equations recovered from the CNF encoding; the other combines exactly-one constraints and conditional equalities to show that incompatible boundary assignments violate count conservation.

\subsection{Public Interface for Programming Strategies}
\label{sec:batch-rationale}

We define a strategy as a program that selects and schedules solver operations based on the current solver state in order to decide the satisfiability of a formula. The front end of \sys provides methods to invoke an operation or a batch thereof, and to observe the result of the invocation. 

\subsubsection{Applying operations.}

The front end allows the user to invoke operations supported by the \sys backend at various granularity. The function \texttt{rules\_\allowbreak applicable} lists operations whose requirements on the current solver state hold: for example, BVE requires decision level zero and decide requires there is no unpropagated unit. The \texttt{configure} method can be used to specify a batch of the same operation (e.g., BVE, vivify, local search) with particular settings and effort budgets. The concrete settings depend on the operation: for BVE, one can set the increase in the number of clauses caused by each elimination and a pre-defined heuristic for prioritizing variables to eliminate. Similarly, vivification allows selection of a tier of learned clauses, and local search accepts a random seed. Effort measures also vary: unit propagation is measured by propagation count, while local search charges work for updating clauses as variables are flipped. During the execution, the user can configure the same operation with different settings and limits. After the \texttt{configure} method is called, the backend validates that this is an applicable operation and registers the request. The \texttt{advance} method executes the currently configured operation repeatedly until either the configured budgets are exhausted or a stopping point is reached (e.g., a conflict is detected). A strategy programmed over the front-end can involve invoking \texttt{configure}, one or more invocations of \texttt{advance}, and another \texttt{configure}. In addition to applying a batch of operations, the front end also allows applying a single step of a certain operation (e.g., eliminating a single variable, deciding on a single variable) with the \texttt{apply\_instance} method, where the meaning of the target is determined by the applied operation. For BVE, the target is the variable to eliminate; for a decision operation, it is a literal.\footnote{The documented public interface in \texttt{include/tactix.h} lists settings, effort limits, target selections, and result types.} 

\subsubsection{Observing operation effects and solver state.}

Each call to \texttt{advance} reports whether the operation made progress, why it stopped, and how much work it performed in a result record. Reasons to stop include completion, exhaustion of the allowance, inapplicability, conflicts, and final SAT or UNSAT answers. The function \texttt{stat} lets a strategy inspect named measurements maintained by the solver. The \texttt{statistics\_id} argument shown in Figure~\ref{fig:architecture} selects which quantity to read, such as the total number of conflicts, variables eliminated, or resolution attempts. It also reports current state, such as the decision level and the number of unassigned variables. Reading an accumulated total before and after an operation gives the amount attributable to that operation. For example, the user-defined strategy can compare variables eliminated with resolution attempts during a batch of BVE operations, then decide whether to request further eliminations, relax its clause-growth bound, or try another technique.\footnote{The \texttt{TACTIX\_STAT\_*} identifiers in \texttt{include/tactix.h} list the available measurements.} The front end also allows one to inspect results after an individual operation. When an effort limit is reached, the back end will finish any transformation in progress, including its proof and model-reconstruction updates, before returning control, so that subsequent operations can proceed from a consistent state.

\subsubsection{Programming a new strategy.}

Using the front-end methods, one can program strategies ranging from complete search procedures to specialized simplification sequences. The sequence of operations allowed by the front end can be represented with the following pattern:
\begin{center}
  \small
  \begin{tabular}{@{}l@{}}
    \texttt{(simplification | structural reasoning | local search)* ;} \\
    \texttt{(search ; restart-to-root ;} \\
    \quad\texttt{(simplification | structural reasoning | local search)*)*}
  \end{tabular}
\end{center}
Here, \texttt{;} denotes sequential composition, \texttt{|} denotes choice, and \texttt{*} denotes zero or more repetitions. In words, \sys allows arbitrary sequences of simplification, structural reasoning, and local search at decision-level zero, interleaved with CDCL search sessions. The same operation can be invoked with different settings and efforts within the same execution. Internally, the solver backend may need to initialize new heuristics required by the selected operations during execution. The default strategy of \sys is programmable top of the public interface. On the other hand, for a specific circuit encoding, a user-defined strategy might alternate congruence reasoning, which discovers equivalent signals, with substitution, which replaces equivalent literals. The strategy could use \texttt{stat} to measure newly discovered equivalences and substituted variables, and repeat the sequence while it remains productive. The user can also fall back to the default strategy by invoking \texttt{default\_\allowbreak solve} after performing the sequence they designed. The back end will continue the solving process and in the end either construct a model to the original formula or produce a uniform UNSAT proof. Section~\ref{sec:evaluation} describes and evaluates two alternative strategies for two specific application domains that resulted in substantial performance gains.

\section{Solver Development Methodology}
\label{sec:harness}
\label{sec:codevelopment}

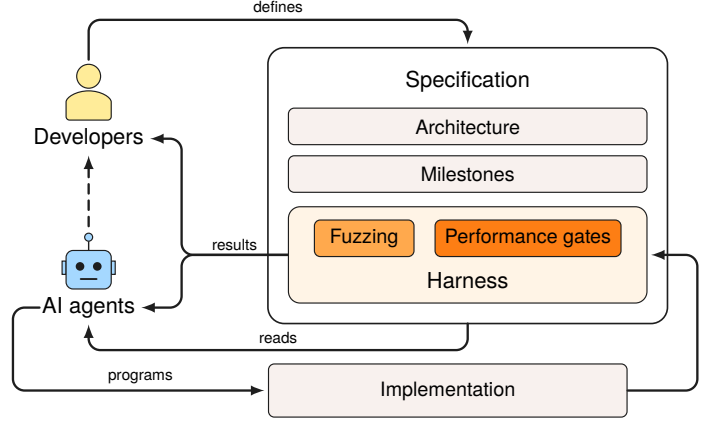
\begin{figure}[t]
  \centering
  \resizebox{.75\linewidth}{!}{\begin{tikzpicture}[
    font=\sffamily,
    outline/.style={draw=ink, line width=0.5pt},
    card/.style={outline, rounded corners=2.5pt, align=center, inner sep=2.5pt, font=\sffamily\footnotesize},
    heading/.style={font=\sffamily\normalsize, inner sep=1pt},
    note/.style={font=\sffamily\scriptsize, inner sep=1pt},
    flow/.style={draw=ink, line width=1pt, line cap=round, line join=round,
                 rounded corners=5pt, -latex, shorten >=1.5pt},
    robot/.pic={
      \draw[ink, line width=0.6pt] (0,0.16) -- (0,0.3);
      \draw[outline, fill=aifill] (0,0.36) circle (0.06);
      \draw[outline, fill=aifill] (-0.4,-0.2) rectangle (-0.34,0);
      \draw[outline, fill=aifill] (0.34,-0.2) rectangle (0.4,0);
      \draw[outline, fill=aifill, rounded corners=3pt] (-0.34,-0.42) rectangle (0.34,0.16);
      \fill[ink] (-0.13,-0.08) circle (0.065) (0.13,-0.08) circle (0.065);
      \draw[ink, line width=0.8pt, line cap=round] (-0.12,-0.27) -- (0.12,-0.27);
    },
    person/.pic={
      \draw[outline, fill=humanfill, rounded corners=1.5pt] (-0.34,-0.42) -- (-0.34,-0.2)
        arc[start angle=180, end angle=0, x radius=0.34, y radius=0.22] -- (0.34,-0.42) -- cycle;
      \draw[outline, fill=humanfill] (0,0.24) circle (0.18);
    },
  ]
  \definecolor{ink}{HTML}{1E1E1E}
  \definecolor{archfill}{HTML}{F8F1EE}
  \definecolor{harnessfill}{HTML}{FFF4E6}
  \definecolor{fuzzfill}{HTML}{FFA94D}
  \definecolor{perffill}{HTML}{FD7E14}
  \definecolor{aifill}{HTML}{A5D8FF}
  \definecolor{humanfill}{HTML}{FFEC99}

  \node[heading] (spectitle) at (0,0) {Specification};
  \node[card, fill=archfill, minimum width=5.4cm, minimum height=0.55cm, anchor=north]
    (arch) at ([yshift=-0.2cm]spectitle.south) {Architecture};
  \node[card, fill=archfill, minimum width=5.4cm, minimum height=0.55cm, anchor=north]
    (miles) at ([yshift=-0.15cm]arch.south) {Milestones};
  \node[outline, rounded corners=5pt, fill=harnessfill, minimum width=5.4cm, minimum height=1.45cm,
        anchor=north] (harness) at ([yshift=-0.2cm]miles.south) {};
  \node[card, fill=fuzzfill, minimum width=1.5cm, minimum height=0.55cm, anchor=east]
    (fuzz) at ([xshift=-0.8cm, yshift=0.25cm]harness.center) {Fuzzing};
  \node[card, fill=perffill, minimum width=2.8cm, minimum height=0.55cm, anchor=west]
    (perf) at ([xshift=0.3cm]fuzz.east) {Performance gates};
  \node[heading, anchor=south] at ([yshift=0.2cm]harness.south) {Harness};

  \coordinate (specnw) at ([xshift=-0.3cm, yshift=0.3cm]arch.west |- spectitle.north);
  \coordinate (specse) at ([xshift=0.3cm, yshift=-0.3cm]harness.south east);
  \draw[outline, line width=0.7pt, rounded corners=7pt] (specnw) rectangle (specse);

  \node[card, fill=archfill, minimum width=5.4cm, minimum height=0.8cm, anchor=north west]
    (impl) at ([yshift=-0.6cm]specnw |- specse) {Implementation};

  \coordinate (colx) at ([xshift=-2.7cm]specnw);
  \node[heading] (devname) at ([yshift=-1.37cm]colx) {Developers};
  \pic[local bounding box=devicon] at ([yshift=0.5cm]devname.north) {person};
  \node[heading] (ainame) at ([yshift=0.24cm]colx |- specse) {AI agents};
  \pic[local bounding box=aiicon] at ([yshift=0.5cm]ainame.north) {robot};
  \draw[flow, dash pattern=on 4pt off 3pt, shorten <=3pt] (aiicon.north) -- (aiicon.north |- devname.south);

  \coordinate (topline) at ([yshift=0.42cm]specnw);
  \draw[flow, shorten <=2pt] (devicon.north) |- (spectitle |- topline) -- (spectitle |- specnw);
  \path (devicon.north |- topline) -- (spectitle |- topline) node[note, midway, above=2pt] {defines};

  \coordinate (splitx) at ([xshift=-1.3cm]specnw);
  \draw[flow] (harness.west) -| (splitx |- devname.east) -- (devname.east);
  \draw[flow] (harness.west) -| (splitx |- ainame.east) -- (ainame.east);
  \node[note, anchor=south east] at ([xshift=-0.12cm, yshift=1pt]specnw |- harness.west) {results};

  \coordinate (readsy) at ([yshift=-0.4cm]specse);
  \draw[flow] (spectitle |- specse) |- (ainame.south |- readsy) -- (ainame.south);
  \path (ainame.south |- readsy) -- (spectitle |- readsy) node[note, midway, above=2pt] {reads};

  \coordinate (progx) at ([xshift=-0.4cm]ainame.west);
  \draw[flow] (ainame.west) -| (progx |- impl.west) -- (impl.west);
  \path (progx |- impl.west) -- (impl.west) node[note, midway, above=2pt] {programs};

  \coordinate (resx) at ([xshift=0.45cm]specse);
  \draw[flow] (impl.east) -| (resx |- harness.east) -- (harness.east);
\end{tikzpicture}}
  \caption{The development workflow of \sys. Correctness and performance tests provide feedback for revising the implementation and its specification.}
  \label{fig:overview}
\end{figure}

In this project, we used AI coding agents (Claude Code and Codex) to help develop \sys. Figure~\ref{fig:overview} summarizes our AI-assisted solver development workflow. The developers begin with a specification covering three aspects: the solver architecture, the performance milestones that guide development (Section~\ref{ss:perfomancetesting}), and a testing harness for evaluating correctness and performance (Sections~\ref{ss:correctnesstesting} and~\ref{ss:perfomancetesting}). 

In practice, the specification is stored as a development plan document that record the system design and the non-negotiable requirements, a progress document that keeps track of the milestones, and a test harness that contains scripts and tools that the AI cannot modify without permission. The agents use the specification to implement and refine the solver. They also have access to reference implementations, including \minisat, \glucose, \kissat, \cryptominisat, \cadical, \lymphosat, and \satsumakissat, to study established techniques, how they are efficiently implemented, and investigate differences in solver behavior. The harness evaluates each candidate implementation and reports the results to both the developers and the agents. Together, they use this feedback to investigate correctness failures and performance bottlenecks. The developers and the AI agents are constantly interacting in this process, where the developers ask questions, provide feedback, and give concrete instructions, and the agents report progress, ask questions, identify bottlenecks, and make suggestions. The developers need to continuously revise, concretize, and amend the specification, as well as inspecting the implementation to ensure their consistency with the specified architectural design.

\subsection{Correctness testing}
\label{ss:correctnesstesting}
The flexibility provided by \sys requires ensuring that its techniques work correctly under different user-defined strategies. A strategy may invoke techniques in orders that do not occur in the default strategy, or repeat a technique with different settings and effort limits. Testing individual techniques is therefore insufficient: we must also test their interactions under different strategies.
To address this challenge, we design an evaluation for both formulas and strategies, following established SAT fuzzing approaches, including \texttt{cnffuzz} and \cadical{}'s \texttt{Mobical}~\cite{brummayer2010fuzzing,biere2024cadical}, and model-based testing~\cite{artho2013modelbased}. The harness contains a script to generate random operation sequences that we expect to support (as describe in Section~\ref{sec:architecture}) by repeatedly invoking operations and varying their settings and effort limits. We also design dedicated stress strategies to exercise techniques that random strategies rarely reach, using counters to confirm that the intended transformations occur. In practice, the random policy fails fuzzing much more often than the default policy. And when a test fails, we reduce both the formula and the operation sequence to obtain a small regression test that preserves the failure. One interesting failure cases occurred when a random strategy requested a decision before unit propagation converged. In this situation, the implementation could assign implied literals at the wrong decision level and later discard them incorrectly during backtracking. The default strategy did not expose this error because it always completed propagation before making a decision. In this case, we instead corrected the applicability check to reject decision requests while propagation is pending.
To validate solver results, the harness independently checks satisfying assignments against the original formula and verifies UNSAT proofs using off-the-shelf external checkers. It is part of the specification that the solver must be able to produce proofs for any refutation operation it implements. The availability of trusted external proof checkers is \emph{crucial} for maintaining the correctness of the implementation and catching soundness issues in time.

\subsection{Performance testing and milestones}
\label{ss:perfomancetesting}
Alongside the separation of strategies from proof rules, achieving state-of-the-art performance with a default strategy is also a non-negotiable development goal. Therefore the harness also conduct performance testing on real benchmarks. The annual SAT Competitions~\cite{satcomp2023proceedings,satcomp2024proceedings,satcomp2025proceedings,satcomp2026} provide benchmark collections and comparative solver evaluations for performance testing. The benchmark must be selected thoughtfully: easy instances may fail to expose bottlenecks, while instances that consistently time out provide little feedback on incremental improvements and prolong the development process. Throughout the development process, we defined six milestones (Table~\ref{tab:milestone-targets}) and designed different tiers of benchmark sets (Table~\ref{tab:tiers}) for performance testing.

\paragraph{Development milestones.}
Table~\ref{tab:milestone-targets} summarizes the performance gates and regression tests used for each milestone. The techniques and the gates might be adjusted during the development process and the table shows the final ones when the developer declares the milestone reached. The regression tests B28+C5 consist of all of the Tier B benchmarks and five of the Tier C benchmarks where the performance tends to be sensitive to implementation changes. Each milestone and its regression tests were set dynamically after the preceding milestone was reached. Section~\ref{sec:development-evaluation} evaluates the archived versions associated with these milestones on SAT Competition benchmarks.

\begin{table}[t]
    \caption{Development milestones, performance gates, and retained
    benchmark sets for regression testing.}
    \label{tab:milestone-targets}
    \centering
    \small
    \setlength{\tabcolsep}{4pt}
    \renewcommand{\arraystretch}{1.15}
    \begin{tabular*}{\linewidth}{@{\extracolsep{\fill}}
      c
      >{\raggedright\arraybackslash}p{.58\linewidth}
      >{\raggedright\arraybackslash}p{.23\linewidth}@{}}
      \toprule
      Milestone & Performance Gate & Regression Tests \\
      \midrule
      1 & Each Tier~A instance solved within \SI{30}{s} & None \\
      \addlinespace
      2 & Half of the Tier~B instances solved within \SI{300}{s} & Tier~A \\
      \addlinespace
      3 & PAR-2 $\leq 1.10\times$ \kissat on Tiers~B and C & Tiers~A and B \\
      \addlinespace
      4 & Each Tier~R instance solved within \SI{1200}{s};
          aggregate runtime $\leq 1.10\times$ \kissat & B28+C5 \\
      \addlinespace
      5 & Each Tier~W instance solved within \SI{1200}{s} & B28+C5 and R \\
      6 & Selected benchmarks with GF(2) solved within \SI{600}{s} & B28+C5, R, W \\
      \bottomrule
    \end{tabular*}
  \end{table}

\begin{table}[t]
  \setlength{\tabcolsep}{6pt}

  \caption{Benchmark tiers used during development.}
  \label{tab:tiers}
  \centering
  \small
  \begin{tabular}{@{}cccr@{}}
    \toprule
    Tier & \# & Use & Source \\
    \midrule
    A & 200 & milestone 1 & simple crafted benchmarks \\
    B & 28 & milestone 2, regression & SC'23-24, took \kissat 5-30 seconds \\
    C & 30 & milestone 3, regression & SC'23-24, took \kissat 5-300 seconds \\
    R & 50 & milestone 4, regression & SC'24-26, hard for \sys, easy for \kissat \\
    W & 100 & milestone 5, regression & SC'24-26, uniquely solved by one baseline \\
    D & 1,166 & final eval. & SC24--26 development milestones \\
    \bottomrule
  \end{tabular}
\end{table}


\section{Evaluation}
\label{sec:evaluation}

In this section, we present an experimental evaluation that aims to answer the following questions. First, we investigate whether \sys can achieve state-of-the-art performance on SAT Competition benchmarks. We will show that the answer is positive. Second, we evaluate how much overhead the public interface incurs by comparing two implementations of the default strategies, one over the public interface, the other one directly over the C++ back end. We found that the overhead in this case is negligible. Third, we test the flexibility of \sys on two specialized sets of benchmarks that are known to be hard for modern SAT solvers. Again, the result is encouraging: we are able to find domain-specific strategies that significantly improve upon state-of-the-art solvers on those benchmarks. All experiments were performed on a cluster equipped with AMD EPYC 7513 processors. Per-instance results for these evaluations and the development versions appear in Appendix~\ref{app:per-instance}.

\subsection{Performance of \tactix on SAT Competition Benchmarks}

We first evaluate the default strategy of \sys on SAT Competition (SC) 2024, 2025, and 2026 main-track benchmarks~\cite{satcomp2024proceedings,satcomp2025proceedings,satcomp2026}. We compare \tactix with competition winners of the sequential track in the past three years--\kissatsc{}, \aekissatmab{}, and \satsumakissat--and two additional sequential solvers--\kissat 4.0.4 and \lymphosat{}. Each solver run was given 4 physical cores (even though each solver spawns one thread), a CPU time limit of 5000 seconds, and \SI{32}{GB} of memory. For an UNSAT answer, we checked the generated certificate using the workflow corresponding to its proof format. For DRAT certificates, we used \texttt{drat-trim}\cite{wetzler2014drat} to generate LRAT proofs, which were then checked by \texttt{cake\_lpr}\cite{tan2021cakelpr}. For DSR certificates, we used \texttt{dsr-trim} to generate LSR proofs, which were then checked by Trestle~\cite{codel2024verified}. For PBP certificates, we used VeriPB~\cite{bogaerts2023certified} to elaborate the proofs into a format checked by \texttt{cake\_pb}. For each certificate, proof elaboration and checking were performed sequentially. Each proof-checking run was given 4 physical cores, a CPU time limit of 8 hours, and \SI{64}{GB} of memory.

\begin{table}[t]
  \caption{Results on the SC'24-26 benchmark sets (400 each) and their union (1166). Solved, SAT, and UNSAT count solver answers; parenthesized ``checked'' columns count independently verified solutions.}
  \label{tab:tierd-results}
  \centering
  \scriptsize
  \setlength{\tabcolsep}{2.4pt}
  \renewcommand{\arraystretch}{0.92}
  \resizebox{\linewidth}{!}{%
  \begin{tabular}{@{}llrrrrrrr@{}}
    \toprule
    Set & Configuration & Solved & \shortstack{Solved\\(checked)} & SAT & UNSAT & \shortstack{UNSAT\\(checked)} & \shortstack{Average runtime on\\checked solves (s)} & \shortstack{Average PAR-2\\(checked) (s)} \\
    \midrule
    \multirow{6}{*}{SC24}
      & \tactix & \textbf{355} & \textbf{350} & 167 & \textbf{188} & \textbf{183} & 331.1 & \textbf{1540} \\
      & \satsumakissat & 346 & 339 & 163 & 183 & 176 & 343.8 & 1816 \\
      & \kissat 4.0.4 & 326 & 324 & 159 & 167 & 165 & 352.4 & 2185 \\
      & \lymphosat & 351 & 328 & \textbf{169} & 182 & 168 & \textbf{269.6} & 2021 \\
      & \kissatsc & 327 & 326 & 158 & 169 & 168 & 371.3 & 2153 \\
      & \aekissatmab & 330 & 325 & 165 & 165 & 160 & 368.6 & 2174 \\
    \midrule
    \multirow{6}{*}{SC25}
      & \tactix & \textbf{368} & \textbf{366} & 173 & \textbf{195} & \textbf{193} & 380.3 & \textbf{1198} \\
      & \satsumakissat & 357 & 347 & 166 & 191 & 181 & 394.2 & 1667 \\
      & \kissat 4.0.4 & 336 & 330 & 168 & 168 & 162 & 490.6 & 2155 \\
      & \lymphosat & 355 & 293 & 172 & 183 & 174 & \textbf{339.9} & 2924 \\
      & \kissatsc & 334 & 328 & 166 & 168 & 162 & 488.9 & 2201 \\
      & \aekissatmab & 341 & 335 & \textbf{176} & 165 & 159 & 460.4 & 2011 \\
    \midrule
    \multirow{6}{*}{SC26}
      & \tactix & \textbf{313} & \textbf{312} & \textbf{158} & \textbf{155} & \textbf{154} & 633.8 & \textbf{2694} \\
      & \satsumakissat & 288 & 286 & 135 & 153 & 151 & 665.9 & 3326 \\
      & \kissat 4.0.4 & 243 & 242 & 131 & 112 & 111 & 687.0 & 4366 \\
      & \lymphosat & 271 & 258 & 155 & 116 & 111 & 701.4 & 4002 \\
      & \kissatsc & 253 & 253 & 140 & 113 & 113 & 788.6 & 4174 \\
      & \aekissatmab & 229 & 228 & 136 & 93 & 92 & \textbf{593.8} & 4638 \\
    \midrule
    \multirow{6}{*}{All}
      & \tactix & \textbf{1006} & \textbf{999} & \textbf{481} & \textbf{525} & \textbf{518} & 441.1 & \textbf{1810} \\
      & \satsumakissat & 965 & 946 & 449 & 516 & 497 & 459.6 & 2260 \\
      & \kissat 4.0.4 & 881 & 872 & 444 & 437 & 428 & 495.0 & 2892 \\
      & \lymphosat & 947 & 852 & 478 & 469 & 442 & \textbf{424.0} & 3003 \\
      & \kissatsc & 890 & 883 & 450 & 440 & 433 & 532.0 & 2830 \\
      & \aekissatmab & 875 & 863 & 462 & 413 & 401 & 461.5 & 2940 \\
    \bottomrule
  \end{tabular}
  }
\end{table}

\begin{figure}[t]
  \centering
  \includegraphics[width=\linewidth]{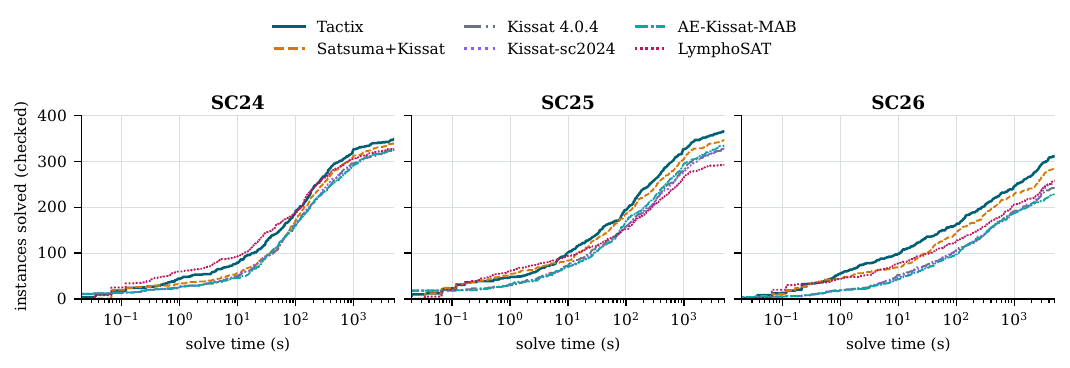}
  \caption{Cumulative checked-solve counts on the three annual competition sets.}
  \label{fig:tierd-cactus}
\end{figure}

\begin{figure}[t]
  \centering
  \includegraphics[width=\linewidth]{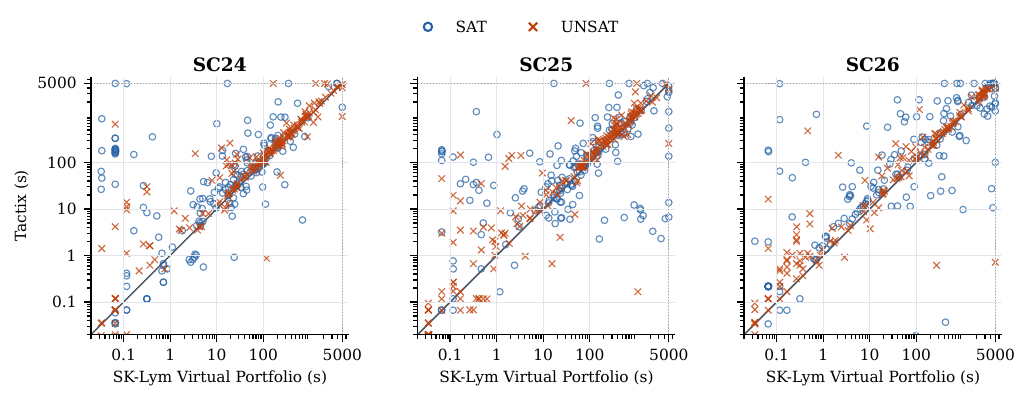}
  \caption{\tactix against a virtual portfolio of \satsumakissat (SK) and \lymphosat (Lym). For every formula, the portfolio uses the faster checked result of the two solvers.}
  \label{fig:tierd-scatter}
\end{figure}

Table~\ref{tab:tierd-results} shows the performance of \tactix across all benchmarks. Figure~\ref{fig:tierd-cactus} shows the cactus plot on the same benchmarks. \tactix outperforms the considered solvers in terms of the number of solved problems, solutions checked by the verifier, and the PAR-2 score, for each benchmark set. Figure~\ref{fig:tierd-scatter} shows a comparison with a virtual portfolio of \satsumakissat and \lymphosat, which covers a large set of SAT-solving techniques. The virtual portfolio in total yields 1002 checked solutions compared to 999 for \tactix. We see that \sys demonstrated complementary performance and contributed a number of unique solutions, including several very quick UNSAT solutions. Upon closer examination, those quick solutions were primarily due to the algebraic reasoning over GF(2), which \lymphosat only partially supports; there is also one instance which \sys solved immediately after the initial symmetry breaking pass but \satsumakissat timed out on. This is due to different criteria for deciding when to perform symmetry breaking.
%
Additional scatter plots appear in Appendix~\ref{app:additional-scatter}.

\begin{figure}[t]
  \centering
  \includegraphics[width=\linewidth]{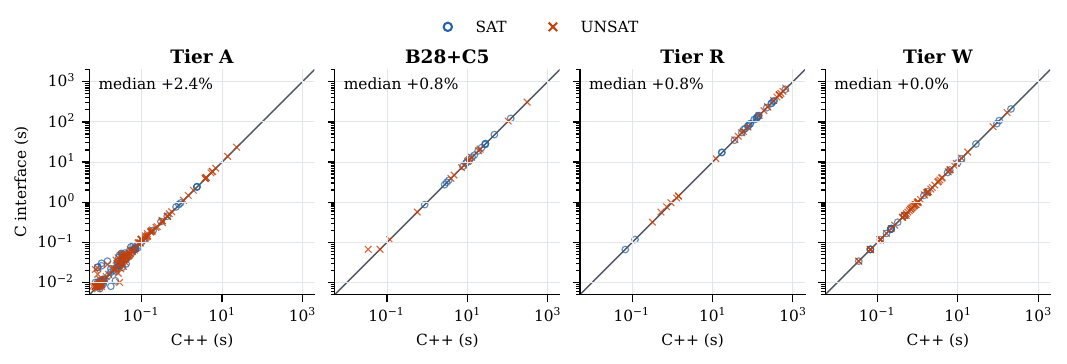}
  \caption{Solving time of the default strategy implemented over the C interface against the same strategy implemented in C++ on four test sets.}
  \label{fig:interface-overhead}
\end{figure}

\subsection{Overhead incurred by the public interface}
\sys comes with two versions of the default strategy: a production version implemented in C++ and is used in the experimental evaluation above, and a client version that is programmed on top of the public C interface. To compare these two versions, we evaluated them on our regular regression test suites Tiers A, B28+C5, R, and W (see Section~\ref{sec:codevelopment}) with a 1{,}200\,s limit. Both configurations solved the same set of instances, with the same observable trace, SAT models, and UNSAT proofs. 

Figure~\ref{fig:interface-overhead} compares their solving times. The median per-instance slowdown of the C interface version is 2.4\% on Tier A, 0.8\% on B28+C5, 0.8\% on R and 0.0\% on W. This suggests that the public interface of \sys incurs negligible overheads when it comes to implementing a complex strategy.

\subsection{Designing domain-specific strategies}

We now demonstrate the flexibility of \sys's strategy interface by designing two domain-specific strategies for two challenging benchmark sets. The first set is a family of formulas that encode the correctness of bit-level multipliers, which is considered challenging for resolution-based solvers~\cite{ritirc2017column,kaufmann2019arithmetic,heisinger2020distributed}. The second set is subproblems from a recent verification effort for Chv\'atal's conjecture~\cite{chvatal_conjecture_1974,looney2026partitioning}, an important open problem in extremal set theory. In both cases, we exploited the structures of the problems to design specialized strategies over the public interface of \sys and show that the resulting strategies can lead to significant performance improvements.

\subsection{Case study 1: arithmetic circuit verification}
\label{sec:crux-strategy}

We first focus on a set of benchmarks on bit-level multiplier verification denoted \texttt{cruxmiter}~\cite{ritirc2017column,kaufmann2019arithmetic}. The benchmark feeds two randomly generated adder trees with the same N one-bit inputs and uses a miter to ask whether their sums can differ. The resulting formula must be unsatisfiable.
We observed that the formulas have a clear circuit structure, with a large number of equivalent signals (due to the two adder trees). Therefore, we instructed an AI agent to design a specialized strategy \tactixmi over the public interface of \sys: after propagating assignments forced at the root, it repeatedly alternates two batches of operations: congruence reasoning recognizes signals that the gate structure makes equivalent, and substitution simplifies the formula by replacing a literal with an equivalent one discovered by congruence reasoning. After each alternation, the strategy checks whether any new equivalences or substitutions were produced. If so, it repeats the process; otherwise, it applies additional inprocessing rules and hands control to a canonical CDCL search loop. In between restarts, the agent decided to only perform clause vivification because it was productive on a test instance, and disabled all other simplification and structural reasoning techniques on this family. The agent was given smaller instances of the same family (N=30) to experiment with different strategies and came up with the above strategy within 5 hours. We then evaluated the resulting strategy on the 10 largest instances of the family (N=32) and compared it with the default strategy of \sys and \kissat 4.0.4. The results are shown in Table~\ref{tab:crux-results}.

Default \sys and \kissat exhibit complementary performance. On the other hand, the specialized strategy is significantly faster than both baselines. Since no symmetry-breaking was triggered, all configurations produced DRAT proof. The proofs generated by \tactixmi are smaller and faster to check than those generated by the other two solvers. On seed 4, the hardest instance, \tactixmi is ~4x faster than \kissat in terms of both solving and proof checking time. We note that it is impossible to implement this strategy in \kissat without modifying its core architecture. This demonstrates that the strategy interface can successfully facilitate the development of novel and practically useful solving strategies.

\begin{table}[t]
  \caption{Performance of a specialized \sys strategy (\tactixmi) on the \texttt{cruxmiter} benchmarks.}
  \label{tab:crux-results}
  \centering
  \tiny
  \setlength{\tabcolsep}{2.1pt}
  \resizebox{0.92\linewidth}{!}{%
  \begin{tabular}{@{}crrrrrrrrrr@{}}
    \toprule
    \multicolumn{1}{c}{Seed} &
      \multicolumn{3}{c}{\tactix} &
      \multicolumn{3}{c}{\tactixmi} &
      \multicolumn{3}{c}{\kissat~4.0.4} \\
    \cmidrule(lr){2-4}\cmidrule(lr){5-7}\cmidrule(lr){8-10}
      & Solve (s) & Check (s) & Proof (GB)
      & Solve (s) & Check (s) & Proof (GB)
      & Solve (s) & Check (s) & Proof (GB) \\
    \midrule
    0 & 2167.6 & 8071.6 & 4.64 & \textbf{823.2} & \textbf{1510.7} & \textbf{1.89} & 883.1 & 1828.0 & 1.96 \\
    1 & 2167.3 & 8216.7 & 4.64 & \textbf{826.9} & \textbf{1515.3} & \textbf{1.89} & 880.6 & 1829.5 & 1.96 \\
    2 & 276.3 & 977.1 & 0.70 & \textbf{223.9} & \textbf{432.2} & \textbf{0.54} & 599.2 & 1284.3 & 1.53 \\
    3 & 1806.1 & 5556.7 & 3.61 & \textbf{277.0} & \textbf{487.2} & \textbf{0.67} & 342.4 & 661.6 & 0.80 \\
    4 & 4373.0 & 13271.7 & 7.97 & \textbf{1386.2} & \textbf{3174.2} & \textbf{3.30} & 6099.0 & 12462.0 & 11.64 \\
    5 & 586.9 & 2052.3 & 1.47 & \textbf{122.0} & \textbf{193.9} & \textbf{0.31} & 741.2 & 1602.5 & 1.71 \\
    6 & 437.9 & 1480.0 & 1.05 & \textbf{283.6} & \textbf{490.5} & \textbf{0.70} & 460.5 & 914.8 & 1.14 \\
    7 & 196.8 & 599.0 & 0.50 & \textbf{41.3} & \textbf{64.1} & \textbf{0.10} & 115.9 & 198.3 & 0.32 \\
    8 & 597.3 & 2060.4 & 1.44 & \textbf{453.6} & \textbf{963.2} & \textbf{1.28} & 819.3 & 1872.3 & 1.97 \\
    9 & 2871.9 & 9527.1 & 5.50 & \textbf{421.8} & \textbf{802.9} & \textbf{0.93} & 5084.3 & 10479.6 & 9.25 \\
    \bottomrule
  \end{tabular}%
  }
\end{table}

\subsection{Case study 2: Chv\'atal's Conjecture}
\label{sec:chvatal-strategy}

\begin{figure}[t]
  \centering
  \includegraphics[width=\linewidth]{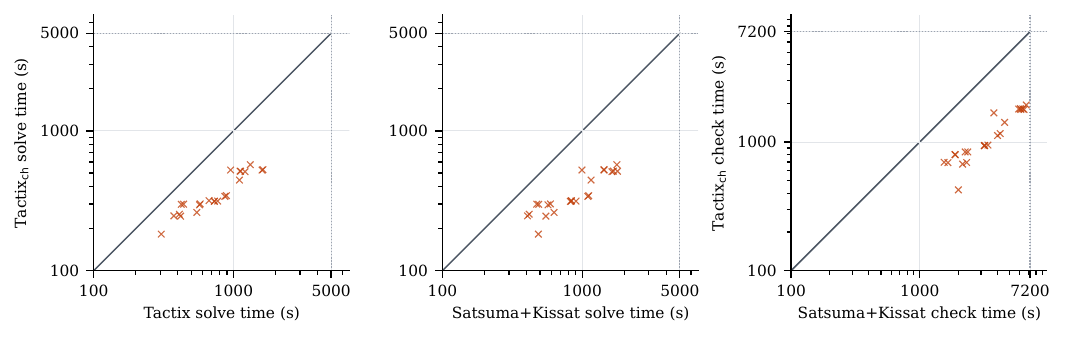}
  \caption{\tactixch  on the 23 Chv\'atal subproblems. Left and middle: solving time against \tactix and \satsumakissat. Right: proof-checking time against \satsumakissat.}
  \label{fig:chvatal-scatter}
\end{figure}

Our second case study focuses on the Chv\'atal's conjecture~\cite{chvatal_conjecture_1974}, a long-standing open problem in extremal combinatorics. Recent work verifies finite cases of this conjecture by partitioning the space of candidate counterexamples into subproblems~\cite{looney2026partitioning}. The subproblems are challenging for modern SAT solvers, as they contain a number of long cardinality constraints~\cite{bailleux_totalizer_2003}. It was reported that solver parameter tuning, preprocessing with \satsuma{}, and solving with cardinality-aware solvers~\cite{klausesclauses} did not significantly help improve performance~\cite{looney2026partitioning}. We consider all of the subproblems for finite case 7 that \kissat~4.0.4 solves in \SI{300}{s} to \SI{1800}{s}. We gave 3 of the 23 subproblems to an AI agent and asked it to design a specialized strategy. Within 2 hours, it found that the totalizer constraints in the encoding count the same variables in the same order, and came up with a strategy, denoted \tactixch, that applies substitution-redundancy steps to merge the duplicated counter variables. This preprocessing merges 4,508 duplicate counter variables, corresponding to 53.1\% of the encoding’s auxiliary variables. After this preprocessing pass, the agent falls back to the default strategy. We evaluated \tactixch on all 23 subproblems and compared it with the default strategy of \sys, \kissat~4.0.4, and \satsumakissat. The results are shown in Table~\ref{tab:chvatal-results} and Figure~\ref{fig:chvatal-scatter}. 

\satsumakissat was not able to identify the symmetry and performed worse than \kissat 4.0.4, which is consistent with what was reported in previous work~\cite{looney2026partitioning}. On the other hand, the variable merging enabled by \tactixch resulted in at least 2x speed up overall against any baselines. This suggests that knowledge of the encoding can guide useful simplifications, and that \sys enables users to implement such simplifications as specialized solving strategies.

\begin{table}[t]
  \caption{Performance of a specialized \sys strategy (\tactixch) on 23 Chv\'atal subproblems.}
  \label{tab:chvatal-results}
  \centering
  \small
  \begin{tabular}{@{}lrrrr@{}}
\toprule
Configuration & Solved & Runtime (s) & Checked & Check time (s)\\
\midrule
\tactixch & 23 & 8,463 & 23 & 26,562 \\
\tactix & 23 & 18,799 & 23 & 41,357 \\
\satsumakissat & 23 & 22,121 & 23 & 85,165\\
\kissat~4.0.4 & 23 & 21,791 & 23 & 34,151 \\
\bottomrule
\end{tabular}

\end{table}

\section{Development progress and discussion}
\label{sec:development-evaluation}

\begin{figure}[t]
\centering
\includegraphics[width=\linewidth]{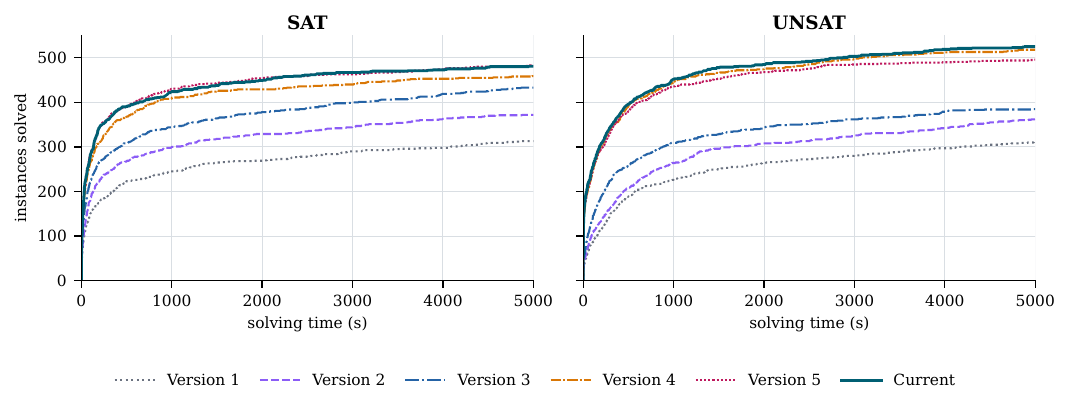}
\caption{Performance of the evaluated versions on SAT and UNSAT benchmarks from SC24--26.}
\label{fig:milestone-ladder}
\end{figure}

In this section, we study our AI-assisted solver development process and discuss what additional infrastructure might be needed to help accelerate it. We identified the source versions associated with the six performance milestones in Table~\ref{tab:milestone-targets}. 
We evaluated how performance improved over the development process by running the five previous versions on the 1,166 SC24--26 benchmarks. Each solver was given the same computational resources: 4 physical cores, 32\,GB of memory, and a 5,000\,s time limit. The cumulative distributions of solving time for SAT and UNSAT instances are shown in Figure~\ref{fig:milestone-ladder}.

We now describe the actual implementation work in each of the versions and relate it to performance. Version~1 implemented basic CDCL search with restarts and learned-clause reduction, which already solves a majority of the benchmarks. Version~2 added inprocessing techniques including BVE, vivification, subsumption, and congruence. Version~3 added local search, SAT sweeping, and blocked and covered clause elimination. Both versions resulted in substantial net gains in solved instances.
The largest increase came in the 37 days from version~3 to version~4. Version~4 improved variable-elimination and
congruence implementations and integrated structured symmetry preprocessing and native substitution-redundancy proofs. This version newly solved 183 formulas and lost 24, for a net gain of 159. Of the newly solved formulas, 134 were UNSAT.
Milestone~5 shifted attention to a fixed set of 100 formulas for which exactly one of the baselines had a checked solve within \SI{1200}{s}. It turned out that most were covered by \satsumakissat or \lymphosat{}. To reach the milestone, version~5 incorporated several structural reasoning techniques from \lymphosat as local search passes and improved the proof-checking pipeline. Reaching the target took nine days. Interestingly, across the full competition benchmark set, this version achieved a net gain of 24 SAT instances and a net loss of 22 UNSAT instances. This decrease in UNSAT performance was due to an attempt to warm-start CDCL search with partial assignments from the newly introduced local search passes, which turned out to hurt performance. The following 13 days added algebraic reasoning from \cryptominisat and additional structural reasoning inspired by \lymphosat{}. These additions recovered a number of UNSAT instances lost at the previous milestone and resulted in the current version.

\begin{table}[t]
\caption{Performance and implementation activity of the evaluated versions. Days are counted from project start. Commits and added/deleted lines cover the interval since the preceding row, or project start.}
\label{tab:development-timeline}
\label{tab:development-activity}
\centering
\footnotesize
\setlength{\tabcolsep}{3pt}
\renewcommand{\arraystretch}{1.15}
\begin{tabular*}{\linewidth}{@{\extracolsep{\fill}}llrrrrr@{}}
  \toprule
  Version & Commit & Day & Solved & Commits & \shortstack{Solver/API lines\\added / deleted} & \shortstack{Test/fuzzing lines\\added / deleted} \\
  \midrule
  1 & \texttt{dcc3e} & 5 & 625  & 47 & 3,679 / 416 & 1,726 / 24 \\
  2 & \texttt{54629} & 11 & 735 & 131 & 9,488 / 1,599 & 3,286 / 225 \\
  3 & \texttt{6f44f} & 34 & 818 & 831 & 48,943 / 24,954 & 15,295 / 3,150 \\
  4 & \texttt{f50e4} & 71 & 977 & 1,003 & 57,939 / 13,778 & 37,196 / 1,782 \\
  5 & \texttt{0a906} & 80 & 979 & 246 & 13,282 / 1,383 & 8,964 / 447 \\
  Now & \texttt{0f7c1} & 93 & 1,006 & 603 & 84,566 / 17,540 & 59,077 / 9,253 \\
  \bottomrule
\end{tabular*}
\end{table}

Table~\ref{tab:development-activity} also records repository activity between versions. Before milestone~3, substantial changes were concentrated in the default strategy, solver core, and simplification techniques. Between milestones 3 and 4, the work extended to symmetry reasoning, configurable strategies, and diagnostic tools for comparing search behavior. The final interval includes extensive structural reasoning and public-interface work, together with 59,077 added and 9,253 deleted lines of test and fuzzing code. It also includes 16,144 added and 1,034 deleted
lines in reusable experiment tools. These changes supported correctness, testing, and programmability alongside performance improvements.

Testing and fuzzing also accounted for a substantial share of implementation activity. Counting both added and deleted lines in Table~\ref{tab:development-activity}, changes to test and fuzzing code amounted to approximately 51\% of changes to solver and API code over the full development period. This ratio varied across intervals, rising from 25\% in the interval leading to version~3 to 54\%, 64\%, and 67\% in the three subsequent intervals. The later stages involved an increasing amount of test and fuzzing code change relative to solver code change. This is because correctness testing becomes more challenging as new techniques and strategy interactions are introduced (Section~\ref{ss:correctnesstesting}). 

\subsubsection{Infrastructure for AI-assisted solver development.}
While we believe the use of AI agents accelerated our solver development process, we still encountered a number of bottlenecks. To ensure the correctness of our implementation, whenever we committed changes that affected solver behavior, we performed extensive fuzzing of both random strategies and the default strategy, in both production and debugging modes, and ran regression tests. In this process, we checked each generated UNSAT proof, which created significant overhead: checking an UNSAT proof for a benchmark solved in 10 minutes sometimes took up to an hour. Moreover, proof certificates can be large and create storage pressure if they were not cleaned up promptly. We believe \emph{online proof checking}~\cite{schreiber2024online} is a promising approach to tackle this bottleneck and accelerate the development cycle.
In addition, fixing the performance gate to a small set of benchmarks risks overfitting to them; this was particularly evident in the transition from milestone~4 to milestone~5. It could be interesting to develop \emph{performance fuzzers}, which might construct synthetic formulas by combining specific structures to stress-test particular techniques. This could enable us to evaluate solver performance in a more targeted manner and potentially uncover performance issues much more efficiently.

\section{Related Work}
\label{sec:related}

Modern SAT solvers provide several ways to influence solving. \cadical offers configuration options, incremental solving, and interfaces for external propagation and proof tracing~\cite{biere2024cadical}. In particular, IPASIR-UP allows an external program to propagate literals, supply reasons, and influence branching decisions~\cite{fazekas2023ipasirup}. These interfaces support close interaction with a running solver. Automated parameter tuning can substantially improve SAT solver performance~\cite{hutter2009paramils}. \sys broadens the space of expressible strategies by allowing clients to select, configure, and interleave search and inprocessing operations during execution. This interface supports experimentation with new solving strategies using the reasoning techniques and proof systems implemented by the backend.
The separation of reasoning rules from their application strategies is well established in abstract SAT and SMT calculi~\cite{nieuwenhuis2006abstract,jarvisalo2012inprocessing}. Mari\'c's formalization, which underlies \argosat{}, separates heuristic choices from the algorithms whose correctness is proved~\cite{maric2009formalization}. \sys builds on this principle by exposing the techniques of a proof-producing inprocessing CDCL solver through a public strategy interface. Strategy languages in SMT solvers, such as Z3, allow users to compose simplification and solving procedures into tactics~\cite{demoura2013strategy}. \sys is designed to expose similar flexibility within a sequential CDCL solver.
\sys uses established proof formats and checking tools. DRAT and LRAT support clausal proof production and efficient checking~\cite{wetzler2014drat,cruzfilipe2017lrat}; substitution redundancy supports transformations such as symmetry breaking~\cite{buss2021sr,codel2024verified}. Our testing infrastructure also builds on formula fuzzing and model-based API testing~\cite{brummayer2010fuzzing,artho2013modelbased}, extending them to varied and randomly generated sequences of solver operations.

\section{Conclusion}
\label{sec:conclusion}
\label{sec:discussion}

We introduced \sys, which is both a high-performing proof-producing CDCL solver and a library that allows users to program novel solving strategies. \sys achieves this by separating the solving techniques from the strategies that orchestrate them. \sys performs competitively against other state-of-the-art solvers on benchmarks from the past three editions of the SAT Competition. In two different SAT-solving applications, we showed that one can use \sys to define solving strategies that deviate significantly from those in existing SAT solvers and lead to substantial performance gains. We also reported on our AI-assisted solver development process and discussed potential directions for infrastructure development in the formal methods community.
In addition to supporting additional operations~\cite{soos2009cryptominisat,shah2025learning}, \sys provides infrastructure for systematically evaluating and developing SAT-solving strategies based on online learning, a vision shared by several existing lines of work~\cite{shi2025dynamicsat,wu2025cubingtuning,ganeshfmsd22}. Finally, it would be interesting to develop a domain-specific language on top of \sys's API to make it more convenient to synthesize domain-specific SAT solvers.

\vspace{-0.2cm}
\subsubsection{AI Use Statement}
Claude Code and OpenAI Codex were used to assist with implementation, testing, debugging, and manuscript preparation, including language editing and plot generation. Section~\ref{sec:harness} describes the human-supervised AI-assisted solver development process. The authors are responsible for the content and conclusions of this paper.

\noindent\textbf{Declaration of Competing Interests}
The authors declare no competing interests.

\newpage
\bibliographystyle{splncs04}
\bibliography{references}

\newpage
\appendix

\newgeometry{margin=10mm}
\section{Additional plots and results}
\label{app:additional-scatter}

\begin{figure}[h]
  \centering
  \includegraphics[width=.64\linewidth]{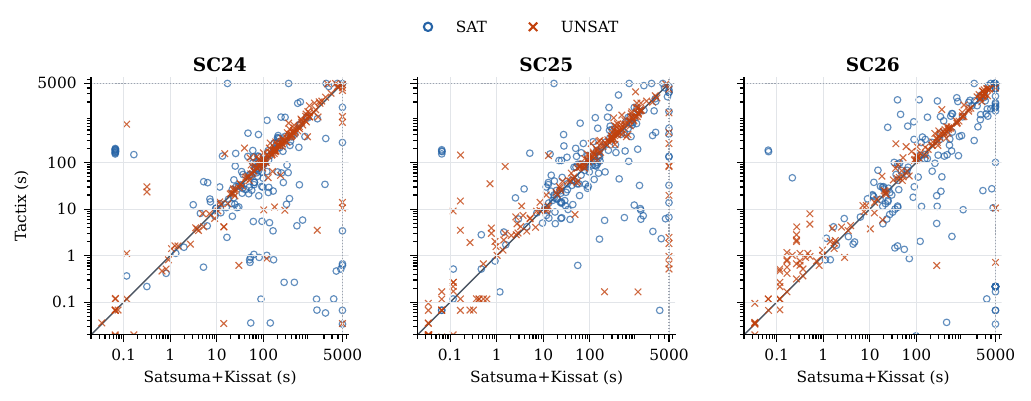}
  \caption{Tactix against \satsumakissat on SC24, SC25, and SC26.}
  \label{fig:tierd-scatter-satsuma}
\end{figure}

\begin{figure}[h]
  \centering
  \includegraphics[width=.64\linewidth]{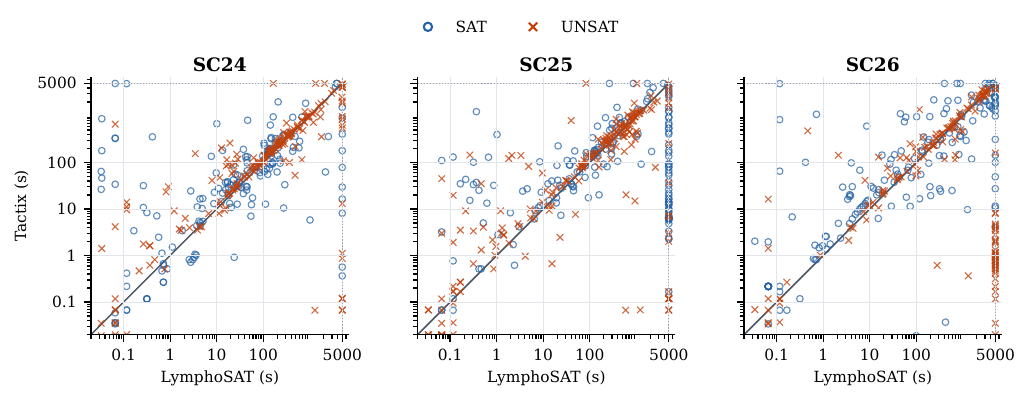}
  \caption{Tactix against \lymphosat on SC24, SC25, and SC26. Circles denote SAT and crosses UNSAT. For a result checked on only one side, the other side is plotted at the \SI{5000}{s} solving-limit boundary; formulas checked by neither are omitted. Points below the diagonal favor Tactix.}
  \label{fig:tierd-scatter-lympho}
\end{figure}

\begin{table}[H]
  \caption{Aggregate \texttt{cruxmiter} results.}
  \label{tab:appendix-crux-checks}
  \centering
  \small
  \begin{tabular}{@{}lrrrr@{}}
    \toprule
    Configuration & Accepted & Solve total (s) & Check total (s) & Proof (GB) \\
    \midrule
    \tactix & 10 & 15,481.213 & 51,812.555 & 31.502 \\
    \tactixmi & 10 & 4,859.624 & 9,634.313 & 11.607 \\
    \kissat 4.0.4 & 10 & 16,025.548 & 33,132.869 & 32.296 \\
    \bottomrule
  \end{tabular}
\end{table}

\section{Software versions}

\paragraph{Evaluated solver versions.}
The Tier~D evaluation uses \tactix commit \texttt{0f7c10174faa}.
The final fuzzing campaign for the evaluated \sys version comprised 10,000 default-strategy cases, 10,000 random-strategy cases, 5,000 cases with graph reasoning disabled, and 5,000 cases with affine reasoning enabled. Each campaign also replayed 51 retained regression cases. No failures were detected and all completed SAT and UNSAT results passed independent model or proof checking. 

\kissat~4.0.4 was built from the \texttt{rel-4.0.4} tag of the
\href{https://github.com/arminbiere/kissat}{upstream repository};
\kissatsc uses commit \texttt{1e56df6c9bd2} of the same repository.
The other baselines were obtained from the official SAT Competition submission archives:
\href{https://satcompetition.github.io/2025/downloads/solvers/main/AE_kissat2025_MAB.tar.xz}{\aekissatmab (2025)},
\href{https://satcompetition.github.io/2026/downloads/solvers/anders.tar.xz}{\satsumakissat (2026)}, and
\href{https://satcompetition.github.io/2026/downloads/solvers/green.tar.xz}{\lymphosat (2026)}.
The SC26 submissions were rebuilt for the cluster's processors without changing their algorithms.

\paragraph{Proof-checker versions.}
The DRAT pipeline uses \texttt{drat-trim} commit \texttt{2e3b2dc0ecf9} and
\texttt{cake\_lpr} commit \texttt{a4323b203cc9}.
The DSR pipeline uses \texttt{dsr-trim} commit \texttt{0b8cc0ef4d22} and Trestle commit \texttt{03ac9418bc23} with an archived bug fix. The PBP pipeline uses VeriPB~3.0.2 and \texttt{cake\_pb} commit \texttt{6edb2f3585f1}.

\clearpage

\section{Per-instance evaluation results}

\label{app:per-instance}

Filenames show the first five hash characters followed by the benchmark name. S/U denotes the reported SAT/UNSAT outcome; no runs disagree on these outcomes. TO denotes a timeout, M a solver memory limit, E1/E134 an exit without an answer, and ? an UNKNOWN result with no recorded termination reason. A dash denotes an unavailable measurement.

\subsection{Competition solvers and development versions}

T denotes \tactix, K4 denotes \kissat~4.0.4, K24 denotes \kissatsc, AE denotes \aekissatmab, SK denotes \satsumakissat, and L denotes \lymphosat. Each competition-solver cell gives solving time / proof-checking time when the latter is available. A superscript * marks an answer not accepted by the checking pipeline; p marks a correct model not accepted under the required output protocol. V1--V5 give solving times for the development versions; these columns do not assert certificate acceptance. Now is identical to T and is shown only once. The virtual SK+L portfolio takes the smaller solving time among accepted results from those two columns. SC lists competition years; instances shared across years are listed once.

\begingroup
\fontsize{6}{6}\selectfont
\setlength{\tabcolsep}{1pt}
\renewcommand{\arraystretch}{.9}
\begin{longtable}{@{}>{\raggedright\arraybackslash}p{.38\linewidth}ll*{11}{r}@{}}
\caption{Per-instance competition and development results (competition cells: solve/check).}\label{tab:instance-competition}\\
\toprule
Benchmark & SC & S/U & T & K4 & K24 & AE & SK & L & V1 & V2 & V3 & V4 & V5 \\
\midrule
\endfirsthead
\toprule
Benchmark & SC & S/U & T & K4 & K24 & AE & SK & L & V1 & V2 & V3 & V4 & V5 \\
\midrule
\endhead
\midrule
\endfoot
\bottomrule
\endlastfoot
0\allowbreak{}0\allowbreak{}d\allowbreak{}1\allowbreak{}f\allowbreak{}-\allowbreak{}l\allowbreak{}e\allowbreak{}c\allowbreak{}\_\allowbreak{}m\allowbreak{}u\allowbreak{}l\allowbreak{}t\allowbreak{}\_\allowbreak{}K\allowbreak{}v\allowbreak{}W\allowbreak{}\_\allowbreak{}1\allowbreak{}2\allowbreak{}x\allowbreak{}1\allowbreak{}1\allowbreak{}.\allowbreak{}s\allowbreak{}a\allowbreak{}n\allowbreak{}i\allowbreak{}t\allowbreak{}i\allowbreak{}z\allowbreak{}e\allowbreak{}d & 24 & U & 4060/10800 & 3900/7550 & 3620/7330 & TO & 3930/6650\textsuperscript{*} & 3430/45000\textsuperscript{*} & ? & ? & ? & 3880 & ? \\
0\allowbreak{}0\allowbreak{}d\allowbreak{}5\allowbreak{}a\allowbreak{}-\allowbreak{}s\allowbreak{}t\allowbreak{}\_\allowbreak{}8\allowbreak{}9\allowbreak{}0\allowbreak{}\_\allowbreak{}8\allowbreak{}6\allowbreak{}\_\allowbreak{}9\allowbreak{}\_\allowbreak{}5\allowbreak{}7\allowbreak{}2\allowbreak{}.\allowbreak{}n\allowbreak{}o\allowbreak{}r\allowbreak{}m\allowbreak{}a\allowbreak{}l\allowbreak{}i\allowbreak{}s\allowbreak{}e\allowbreak{}d & 25 & ? & TO & TO & TO & TO & TO & TO & ? & ? & ? & ? & ? \\
0\allowbreak{}0\allowbreak{}f\allowbreak{}d\allowbreak{}8\allowbreak{}-\allowbreak{}2\allowbreak{}.\allowbreak{}n\allowbreak{}o\allowbreak{}r\allowbreak{}m\allowbreak{}a\allowbreak{}l\allowbreak{}i\allowbreak{}s\allowbreak{}e\allowbreak{}d & 25 & S & 26.3 & 12.1 & 11.9 & 11.8 & 33 & 5.57\textsuperscript{p} & 1300 & 33.1 & 14.5 & 12.1 & 23.9 \\
0\allowbreak{}1\allowbreak{}1\allowbreak{}9\allowbreak{}8\allowbreak{}-\allowbreak{}a\allowbreak{}r\allowbreak{}l\allowbreak{}e\allowbreak{}s\allowbreak{}\_\allowbreak{}t\allowbreak{}h\allowbreak{}r\allowbreak{}e\allowbreak{}s\allowbreak{}2\allowbreak{}0\allowbreak{}\_\allowbreak{}p\allowbreak{}1\allowbreak{}0\allowbreak{}\_\allowbreak{}r\allowbreak{}7\allowbreak{}5\allowbreak{}3\allowbreak{}3 & 26 & U & 0.069/0.14 & 0.01/0.084 & 0.01/0.084 & 0.01/0.084 & 0.117/0.096 & 0.034/0.026 & 0.018 & 0.217 & 0.018 & 0.034 & 0.034 \\
\end{longtable}
\endgroup

\subsection{Public-interface overhead}

C++ and C compare the production implementation with the public C interface. All 383 pairs are included, with shared outcomes for the two implementations. A, B+C, R, and W identify Tier A, B28+C5, Tier R, and Tier W.

\begingroup
\fontsize{6}{6}\selectfont
\setlength{\tabcolsep}{1pt}
\renewcommand{\arraystretch}{.9}
\begin{longtable}{@{}l>{\raggedright\arraybackslash}p{.32\linewidth}lrr@{\hspace{6pt}}l>{\raggedright\arraybackslash}p{.32\linewidth}lrr@{}}
\caption{Per-instance interface runtimes (read down each half).}\label{tab:instance-interface}\\
\toprule
Tier & Benchmark & S/U & C++ & C & Tier & Benchmark & S/U & C++ & C \\
\midrule
\endfirsthead
\toprule
Tier & Benchmark & S/U & C++ & C & Tier & Benchmark & S/U & C++ & C \\
\midrule
\endhead
\midrule
\endfoot
\bottomrule
\endlastfoot
A & c\allowbreak{}o\allowbreak{}l\allowbreak{}o\allowbreak{}r\allowbreak{}-\allowbreak{}n\allowbreak{}1\allowbreak{}4\allowbreak{}-\allowbreak{}k\allowbreak{}3\allowbreak{}-\allowbreak{}s\allowbreak{}0\allowbreak{}0\allowbreak{}6\allowbreak{}.\allowbreak{}c\allowbreak{}n\allowbreak{}f & U & 0.009 & 0.008 & A & x\allowbreak{}o\allowbreak{}r\allowbreak{}-\allowbreak{}n\allowbreak{}5\allowbreak{}6\allowbreak{}-\allowbreak{}s\allowbreak{}0\allowbreak{}0\allowbreak{}1\allowbreak{}.\allowbreak{}c\allowbreak{}n\allowbreak{}f & U & 0.007 & 0.008 \\
A & c\allowbreak{}o\allowbreak{}l\allowbreak{}o\allowbreak{}r\allowbreak{}-\allowbreak{}n\allowbreak{}1\allowbreak{}5\allowbreak{}-\allowbreak{}k\allowbreak{}4\allowbreak{}-\allowbreak{}s\allowbreak{}0\allowbreak{}0\allowbreak{}7\allowbreak{}.\allowbreak{}c\allowbreak{}n\allowbreak{}f & U & 0.008 & 0.008 & A & x\allowbreak{}o\allowbreak{}r\allowbreak{}-\allowbreak{}n\allowbreak{}6\allowbreak{}5\allowbreak{}-\allowbreak{}s\allowbreak{}0\allowbreak{}0\allowbreak{}5\allowbreak{}.\allowbreak{}c\allowbreak{}n\allowbreak{}f & U & 0.007 & 0.008 \\
A & c\allowbreak{}o\allowbreak{}l\allowbreak{}o\allowbreak{}r\allowbreak{}-\allowbreak{}n\allowbreak{}1\allowbreak{}6\allowbreak{}-\allowbreak{}k\allowbreak{}4\allowbreak{}-\allowbreak{}s\allowbreak{}0\allowbreak{}0\allowbreak{}9\allowbreak{}.\allowbreak{}c\allowbreak{}n\allowbreak{}f & U & 0.008 & 0.008 & A & x\allowbreak{}o\allowbreak{}r\allowbreak{}-\allowbreak{}n\allowbreak{}7\allowbreak{}0\allowbreak{}-\allowbreak{}s\allowbreak{}0\allowbreak{}0\allowbreak{}7\allowbreak{}.\allowbreak{}c\allowbreak{}n\allowbreak{}f & U & 0.007 & 0.021 \\
\end{longtable}
\endgroup

\subsection{Case studies}

The seed identifies each of the ten \texttt{cruxmiter} instances ($N=32$); the node identifies each of the 23 Chv\'atal subproblems (finite case 7). All runs returned UNSAT with accepted certificates. Each cell lists solving time / checking time / proof size (GB).

\begin{table}[H]

\begin{minipage}[t]{.43\linewidth}

\caption{Arithmetic verification.}\label{tab:instance-crux}
\centering\fontsize{6}{6}\selectfont\setlength{\tabcolsep}{1pt}
\renewcommand{\arraystretch}{.9}
\begin{tabular}{@{}lrrr@{}}
\toprule
Seed & \tactix & \tactixmi & \kissat~4.0.4 \\
\midrule
0 & 2170/8070/4.64 & 823/1510/1.89 & 883/1830/1.96 \\
1 & 2170/8220/4.64 & 827/1520/1.89 & 881/1830/1.96 \\
2 & 276/977/0.698 & 224/432/0.54 & 599/1280/1.53 \\
3 & 1810/5560/3.61 & 277/487/0.67 & 342/662/0.802 \\
4 & 4370/13300/7.97 & 1390/3170/3.3 & 6100/12500/11.6 \\
5 & 587/2050/1.47 & 122/194/0.311 & 741/1600/1.71 \\
6 & 438/1480/1.05 & 284/491/0.703 & 460/915/1.14 \\
7 & 197/599/0.499 & 41.3/64.1/0.102 & 116/198/0.324 \\
8 & 597/2060/1.44 & 454/963/1.28 & 819/1870/1.97 \\
9 & 2870/9530/5.5 & 422/803/0.931 & 5080/10500/9.25 \\
\bottomrule
\end{tabular}

\end{minipage}\hfill\begin{minipage}[t]{.55\linewidth}

\caption{Chv\'atal's conjecture.}\label{tab:instance-chvatal}
\centering\fontsize{6}{6}\selectfont\setlength{\tabcolsep}{1pt}
\renewcommand{\arraystretch}{.9}
\begin{tabular}{@{}lrrrr@{}}
\toprule
Node & \tactixch & \tactix & \satsumakissat & \kissat~4.0.4 \\
\midrule
1111001 & 525/1800/0.759 & 1600/3760/2.09 & 1430/5880/1.97 & 1620/2770/2.15 \\
11110011 & 528/1820/0.759 & 1630/3640/2.09 & 1430/6060/1.97 & 1620/2770/2.15 \\
111100110 & 344/1160/0.477 & 891/2090/1.17 & 1110/4220/1.49 & 1150/2020/1.56 \\
1111001100 & 340/1120/0.477 & 869/1970/1.17 & 1100/4030/1.49 & 1150/1840/1.56 \\
111111001000 & 511/1790/0.698 & 1210/2500/1.6 & 1630/6110/2.07 & 1680/2580/2.24 \\
1111110010001 & 515/1790/0.698 & 1130/2390/1.52 & 1790/6470/2.13 & 1200/1930/1.64 \\
11111100100010 & 515/1810/0.698 & 1120/2230/1.52 & 1670/6320/2.13 & 1200/1850/1.64 \\
\bottomrule
\end{tabular}

\end{minipage}

\end{table}

\restoregeometry

\end{document}